\documentstyle[preprint,prl,aps,epsf]{revtex}
\large
\begin{document}
\draft
\title
      {
       Relaxation processes in harmonic glasses?
      }
\author{
        G.~Ruocco$^{1}$,
	F.~Sette$^{2}$,
        R.~Di Leonardo$^{1}$,
	G.~Monaco$^{2}$,\\
        M.~Sampoli$^{3}$,
        T.~Scopigno$^{4}$ and
        G.~Viliani$^{4}$
       }
\address{
         $^1$
         INFM and Dipartimento di Fisica, Universit\'a di L'Aquila,
	 I-67100, L'Aquila, Italy. \\
         $^2$
         European Synchrotron Radiation Facility, 
	 B.P. 220, F-38043 Grenoble Cedex, France. \\
         $^3$
         INFM and Dipartimento di Energetica, Universit\'a di Firenze,
	 I-50139, Firenze, Italy. \\
         $^3$
         INFM and Dipartimento di Fisica, Universit\'a di Trento,
	 I-3805, Povo, Trento, Italy. 
        }

\date{\today}
\maketitle
\begin{abstract}
A relaxation process, with the associated phenomenology 
of sound attenuation and sound velocity dispersion, is found 
in a simulated {\it harmonic} Lennard-Jones glass. We propose to 
identify this process with the so called {\it microscopic} 
(or {\it instantaneous}) relaxation process observed in real 
glasses and supercooled liquids. A model based on the memory 
function approach accounts for the observation, and allows to 
relate to each others: 1) the characteristic time and strength 
of this process, 2) the low frequency limit of the dynamic structure 
factor of the glass, and 3) the high frequency sound attenuation 
coefficient, with its observed quadratic dependence on the momentum 
transfer.
\end{abstract}
\vskip 1cm

In a recent paper W.~G\"otze and M.R.~Mayr \cite{Goetze} adapted 
the Mode Coupling Theory \cite{MCT} to glassy phase. This theory, as 
shown numerically for a hard-spheres model, accounts for many of the 
features found in the dynamic structure factor, $S(Q,\omega)$, 
of glasses. Among these: {\it i)} The existence of propagating 
excitations (sound waves), with an almost linear momentum transfer 
($Q$) dependence of their excitation energy $\Omega(Q)$, up to $Q$ 
values that are a significant fraction of $Q_o$, the maximum of the 
static structure factor $S(Q)$. {\it ii)} The quadratic dependence 
on $Q$ of the excitations broadening  $\Gamma(Q)$, and therefore of 
sound attenuation. {\it iii)} The temperature insensitivity of 
$\Gamma(Q)$. {\it iv)} The development in the $S(Q,\omega)$ at 
large $Q$ ($Q/Q_o$$\approx$0.3) of a secondary excitation band, 
at frequencies below the Brillouin peak. The theory also predicts 
two features not yet experimentally detected: {\it a)} A positive 
dispersion of the sound velocity and {\it b)} an intensity "gap" 
in the low frequency region of the $S(Q,\omega)$. In spite of 
the success of this approach, it is still of great interest to 
investigate the physical origin of these phenomena, and, in 
particular, whether they are related to the topological disorder 
and/or to the anharmonicity of the interatomic potential.                 %

In this Letter we report a Molecular Dynamics (MD) simulation 
study of the $Q$ dependence of the sound velocity in a model 
monatomic Lennard-Jones glass in the {\it harmonic} approximation. 
We show that even in a {\it harmonic} glass, by increasing $Q$
there is a positive dispersion of the sound velocity, thus proving
one of the prediction of the MCT theory \cite{Goetze}, and 
relating this phenomenon to the topological disorder. Since 
this dispersion is similar to that found in presence of a relaxation 
process, we attempt to apply a generalized Langevin equation with an 
effective memory function approach to describe the density fluctuations
dynamics. This formalism allows to account for the ubiquitous 
$Q^2$-dependence of the high frequency sound absorption
observed in many glasses by experimenys \cite{Q2g} and by MD 
simulations \cite{Q2MD,dellanna2}.
These results suggest to identify this process with the one
referred to as {\it microscopic} or {\it instantaneous} 
relaxation in {\it real} systems.

The investigated systems consist of $N$=2048, 10976 and 32000 
argon atoms interacting via a (6-12) Lennard-Jones potential 
($\epsilon$=125.2 K, $\sigma$=3.405 A). A standard microcanonical 
MD simulation, performed at decreasing temperatures in the normal 
liquid phase, is followed by a fast quench ($\dot T$$\approx$ 
10$^{12}$ K/s) of the slightly supercooled liquid down to 
$\approx$5 K \cite{Mazz}. Starting from the glass configuration 
at $T$=5 K, the atomic trajectories $\bar r_i(t)$ 
(here $i$=$1...N$ is the particle label) are followed and 
stored for subsequent analysis, and the "inherent" 
configuration  $\{\bar x_i\}_{i=1..N}$ \cite{StWe} at $T$=0 K 
is calculated by the steepest descent method. Two different 
procedures have been used to derive the $S(Q,\omega)$:
(i) The trajectories calculated from MD in the glass 
are used to compute the corresponding time correlation 
function and (ii) The Normal Mode Analysis (NMA) is applied 
to the inherent configuration. This last procedure works in the 
harmonic approximation, which is obtained by retaining only the 
quadratic term of the interaction potential. From the 
calculation of the dynamical matrix ${\bf D}$, one computes 
the eigenvalues, $\omega_p$ ($p$=$1..3N$ is the mode index) 
and the eigenvectors ($\bar e_p(i)$). For large size samples 
(where a direct diagonalization of ${\bf D}$ is not feasible) 
the $S(Q,\omega)$ is obtained by the method of moments \cite{mom}.

In the MD runs, the dynamics structure factor is
calculated from the numerical time-Fourier transform
of the intermediate scattering function $F(Q,t)$,
defined as:
\begin{equation}
\label{fqt}
F(Q,t)= 1/N \langle 
\Sigma_{ij} \exp{(i \bar Q \cdot \bar r_i(t) )}
\exp{(-i \bar Q \cdot \bar r_j(0))}
\rangle .
\end{equation}
In the harmonic framework and in the classical limit, 
the one excitation approximation of the dynamic structure 
factor $S^{^{(1)}}\!(Q,\omega)$ is obtained
by the normal mode expansion of the atomic displacements:
\begin{equation}
\label{udt}
\bar r_i(t)=\bar x_i+
\sqrt{K_BT/M} \Sigma_p \bar e_p(i) A_p(t)/\sqrt{\omega_p}.
\end{equation}
$A_p(t)$ is the amplitude of the $p$-th normal mode 
and is characterized by $\langle|A_p(t)|^2\rangle=1$. 
This gives:
\begin{equation}
\label{sqw}
S^{^{(1)}}\!(Q,\omega)=
(K_BTQ^2/M\omega^2) \; \Sigma_p E_p(Q) \delta(\omega-\omega_p),
\end{equation}
where we have introduced the spatial power spectrum of the 
longitudinal component of the eigenvectors, $E_p(Q)$:
\begin{equation}
\label{eqw}
E_p(Q)=|\Sigma_i (\hat Q \cdot \bar e_p(i))
\exp{{(i \bar Q \cdot \bar x_i )} |^2}.
\end{equation}
Here, $\hat Q$=$\bar Q/|Q|$ and the Debye-Waller factor
has been neglected. Differently from the second 
($\int \omega^2 S(Q,\omega) d\omega$=
$\int \omega^2 S^{^{(1)}}\!(Q,\omega) d\omega$=%
$K_BTQ^2/M$) and higher moments sum rules,
the zeroth moment sum rule for $S(Q,\omega)$, 
$\int S(Q,\omega) d\omega$=$S(Q)$, does not hold for
$S^{^{(1)}}\!(Q,\omega)$, as in this function the 
elastic intensity is missing. Rather, it is useful 
to define the "inelastic" contribution to $S(Q)$:
$S_{_{i}}(Q)$=$\int S^{^{(1)}}\!(Q,\omega) d\omega$.

Selected examples of the $S(Q,\omega)$ calculated for
different size systems, and with different methods are
reported in Fig.~1. As there is a trivial dependence
of the $S^{^{(1)}}\!(Q,\omega)$ on $T$, here we report
this quantity multiplied by the factor $(M/K_BT)$. 
Figure~1a shows the $S(Q,\omega)$ 
in the low $Q$ range, as calculated from the MD
runs for the $N$=32000 particles system, while Fig.~1b
shows the intermediate $Q$ range, in the case of the
harmonic approximation for the $N$=2048 system.
As a check of consistency, the 
inset of Fig.~1b  shows the $Q$ dependence of the 
second moment of $(M/K_BT)S(Q,\omega)$ (+) which, 
according to the sum rule, should be $Q^2$ (full line). 
The inset of Fig.~1a shows, at two $Q$ values, the
comparison of the $S(Q,\omega)$ ($N$=2048) calculated
either from the MD runs (full line) or in the harmonic
approximation ($\circ$). The $S(Q,\omega)$, as derived 
with the two different methods, are equivalent in the 
whole considered $Q$ range. This indicates that the 
Newtonian dynamics at $T$=5 K is truly harmonic, and 
that the results obtained by MD at this $T$ and NMA 
can be interchanged among each other. 

The general features of the $S(Q,\omega)$ reported in 
Fig.~1 are: {\it i)} A Brillouin peak, dispersing and 
becoming broader with increasing $Q$, dominates the
spectrum up to $Q$$\approx$10 nm$^{-1}$ 
({$Q/Q_o$$\approx$0.45). {\it ii)} At larger $Q$ 
values a second peak, observed at frequencies below
the Brillouin peak, starts to dominate the data. 
This secondary peak has been already detected 
in many systems: 1) in liquid water, both experimentally 
\cite{wat97} and by MD \cite{watmix}, 
where it has been interpreted as a signature of the 
transverse dynamics; 2) in vitreous silica, by MD,
where it has been interpreted either 
in terms of a transverse dynamics \cite{dellanna2} or
as an evidence of the Boson peak \cite{kob}; and 3) in
a hard sphere glass, where it has been theoretically 
predicted \cite{Goetze}. In this respect, it is
worth to underline the striking similarity 
between the $S(Q,\omega)$ reported here and those
of Fig.~5 of Ref.~\cite{Goetze}. Finally,
{\it iii)}, the $S(Q,\omega)$ up to 
$Q$$\approx$5 nm$^{-1}$ shows a nearly constant 
and $Q$-independent intensity below the Brillouin 
peak frequency. This plateau, whose value is 
$\lim_{\omega\rightarrow 0} (M/K_BT)S(Q,\omega)$=
$A_0$$\approx$0.03~10$^{-18}$ s$^3$/m$^2$, 
can be identified with the plateau observed in the 
constant $\omega$ cuts of the $S(Q,\omega)$
at $Q$ larger than the Brillouin peak in simulated
\cite{and_v}, calculated (Fig~18 in Ref. \cite{Goetze})
and measured \cite{nh3} systems.

Figure~2 shows the position of the maxima 
($\Omega_{_C}(Q)$) of the current spectra $C_L(Q,\omega)$ 
(=$\omega^2/Q^2S(Q,\omega)$) as a function of $Q$ in the 
investigated $Q$ range. 
The inset of Fig.~2 shows the peak position $\Omega(Q)$ and 
the broadening $\Gamma(Q)$ of the $S(Q,\omega)$ in the low $Q$ 
region, as measured directly from the spectra of Fig.~1a. The $Q$ 
dependence of these parameters agrees with the behavior 
observed in all the glasses investigated so far \cite{Q2g}:
a linear (square) dependece of $\Omega(Q)$ ($\Gamma(Q)$).
The behavior of $\Omega_{_C}(Q)$ closely resembles that of 
the acoustic phonon branches in crystals, i.~e.
the almost linear behavior in the small $Q$ region, a 
maximum around $Q_o/2$, and a minimum around $Q_o$. Most
importantly one can clearly detect a positive dispersion 
of the sound velocity, as highlighted by the dashed line 
in Fig.~2. This dispersion is better seen 
in Fig.~3, where the apparent sound velocity, 
$v(Q)$=$\Omega_{_C}(Q)/Q$, is reported. The quantity $v(Q)$
(full dots) undergoes a transition from the "low frequency" 
\cite{nota} sound velocity $v_o(Q)$ towards the infinite frequency 
sound velocity $v_{\infty}(Q)$.
Here $v_o(Q)$ is calculated as 
$v_o(Q)$=$\sqrt{K_BT/MS_i(Q)}$ (dashed line),
and $v_\infty(Q)$ is calculated either as the
fourth moment of the calculated $S(Q,\omega)$ (open points)
or from the expression of the fourth moment in terms of both
the pair correlation function and the interaction potential 
(full line) \cite{baluc}.
The velocity dispersion is observed up to $Q$$\approx$5 nm$^{-1}$, 
i.~e. $\Omega_{_C}$$\approx$35 cm$^{-1}$, and an opposite
dispersion is observed in the region approaching $Q_o$.
These observations recall of a typicall relaxation scenario.
When a relaxation process with characteristic time $\tau$ is 
active in the system, the transition from $v_o$ to $v_{\infty}$ 
takes place when the condition $\omega\tau$=1 is fulfilled. In 
the present case, considering that the first transition is at at 
$Q$$\approx$2 nm$^{-1}$  where $\Omega_{_C}$$\approx$15 cm$^{-1}$, 
the value of $\tau$ results to be around 0.3 ps. The second 
and third transitions between $v_o$ and $v_\infty$ are observed
just below and above $Q_o$ as a consequence of the slowing down 
of the dynamics (deGennes narrowing) around $Q_o$.
Consequently the whole behavior of $v(Q)$, as reported in Fig.~3, 
can be qualitatively understood in terms of a relaxation process 
with a characteristic time of $\tau$$\approx$0.3 ps. 

The identification of a such a {\it relaxation process} in an 
{\it harmonic system}, suggests to use the formalism that 
describes the density correlators $\phi(Q,t)$=$F(Q,t)/S(Q)$ 
through its generalized Langevin equation \cite{baluc}:
\begin{equation}
\label{lang}
\ddot \phi(Q,t)+\omega^2_o \phi(Q,t)+
\mbox{$\int$}_o^t m(Q,t-t') \dot \phi(Q,t') dt =0
\end{equation}
where $\omega^2_o$=$K_BTQ^2/MS(Q)$ and $m(Q,t)$ is the 
"memory function". This 
equation has been rigorously derived for ergodic 
systems, but, as recently shown in the framework of the MCT
\cite{Goetze}, 
it can be still applied in the non-ergodic glassy phase 
making the the following substitutions:
$\phi(Q,t)\rightarrow \phi'(Q,t)$=$(\phi(Q,t)-f_{_Q})/(1-f_{_Q})$, 
$\omega_o^2(Q)\rightarrow \omega^2_{o \mu}$
$\equiv$$\omega^2_{\infty \alpha}$=$K_BTQ^2/MS_i(Q)$
and $m(Q,t)\rightarrow m_\mu(Q,t)$=$m(Q,t)\!-\!\omega^2_o 
f_{_Q}/(1\!-\!f_{_Q})$. Therefore, the "vibrational" dynamics
of interest is now described by the correlators $\phi'(Q,t)$, 
which is obtained subtracting from $\phi(Q,t)$ the long time 
plateau level, whose value is the non-ergodicity parameter 
$f_{_Q}$. It is worth to note the subtraction of the constant 
term $S(Q)f_{_Q}$ from F(Q,t) is equivalent to neglect the 
elastic contribution in the $S(Q,\omega)$, and that 
$F'(Q,t)$=$S_i(Q)\phi'(Q,t)$ is the Fourier transform 
of the $S^{^{(1)}}\!(Q,\omega)$ as defined in Eq.~(\ref{sqw}).
The whole dynamic behavior of $\phi'(Q,t)$ is now contained in the
{\it microscopic} contribution to the memory function $m_\mu(Q,t)$.
In the case of a harmonic system, the function $m_\mu(Q,t)$ can be 
explicitly calculated from the eigenvalues and eigenvectors of the 
system. Indeed, the Laplace transform (indicated by hats) of 
Eq.~(\ref{lang}) (after the previously indicated substitutions) 
and a straightforward algebra gives:
\begin{equation}
\label{langs}
\hat m_\mu(Q,\!s)\!=\!\left[ 
\hat \phi'(Q,s)[s^2+\omega^2_{o \mu}]\!-\!s \right]
\! \left[ 1\!-\!s\hat \phi'(Q,s)\right ]^{-1}.
\end{equation}
Then, from an inverse Fourier and a subsequent Laplace transform of 
Eq.~(\ref{sqw}), it is easy to get an explicit expression for 
$\hat \phi'(Q,s)$ to be inserted in Eq.~(\ref{langs}). This gives:
\begin{equation}
\label{ms}
\hat m_\mu(Q,\!s)\!=\!\!\left[ {\Sigma} \frac{E_p(Q)}
{\omega_p^2} \frac{s}{s^2\!\!+\!\omega^2_p}\right ] \!
\left[ {\Sigma} \frac{E_p(Q)}{\omega_p^2} {\Sigma}
\frac{E_p(Q)}{s^2\!\!+\!\omega^2_p}\right ]^{-1}
\!\!\!\!\!\!\!-\!s
\end{equation}
This equation provides an explicit expression of the memory function 
in terms of the system eigenstates.
Considering that $m_\mu(Q,\! t)$ is mainly characterized by parameters 
as its initial value $\Delta_{_Q}^2$ and total area $\Gamma_{_Q}$ and 
therefore by 
a decaying time-scale $\tau_{_Q}$$\approx$$\Gamma_{_Q}/\Delta_{_Q}^2$
\cite{notatau}, one can show that
$\hat m(Q,s$$\rightarrow$$\infty)$=$\Delta_{_Q}^2/s$ and
$\hat m(Q,s$$\rightarrow$0)=$\Gamma_{_Q}$  \cite{baluc}. 
Inserting these limiting values into Eq.~(\ref{ms}) one obtains:
\begin{eqnarray}
\label{limit1}
\Delta_{_Q}^2&=&\left [ \Sigma_p E_p(Q)\omega^2_p \right ]-
\left [ \Sigma_p E_p(Q)\omega^{-2}_p \right ]^{-1}\\
\label{limit2}
\Gamma_{_Q}&=&\left[\Sigma_p \frac{E_p(Q)}
{\omega^2_p} \right]^{-2} \!\!\!
\lim_{s\rightarrow 0}  
\left[ \Sigma_p \frac{E_p(Q)}{\omega_p^2}
\frac{s}{s^2\!+\!\omega^2_p}\right ]
\end{eqnarray}
It is now easy to identify, through the explicit expression of the 
zeroth and fourth moments of Eq.~(\ref{sqw}) that 
$\Sigma_p E_p(Q)\omega^2_p$=$v^2_\infty Q^2$ and
$(\Sigma_p E_p(Q)\omega^{-2}_p)^{-1}$=$v^2_o Q^2$, confirming
that $\Delta_{_Q}^2$=$(v_\infty^2\!-\!v_o^2)Q^2$. 
The determination of $\Gamma_{_Q}$ which, being 
the area of the memory function, coincides with the Brillouin 
broadening in the $\omega\tau$$<<$1 limit is slightly more
involved. Using the representation 
$\delta(x)=1/\pi \; \lim_{s\rightarrow 0} s/(s^2+x^2)$,
in Eq.~(\ref{sqw}) one sees that:
\begin{equation}
\label{sss}
S^{^{(1)}}\!(Q,\omega)\!=\!\frac{K_BTQ^2}{\pi M} 
\lim_{s\rightarrow 0} \Sigma_p \frac{E_p(Q)}
{\omega_p^2} \frac{s}{s^2\!+\!(\omega\!-\!\omega_p)^2}, 
\end{equation}
and comparing of Eqs.~(\ref{sss}) and (\ref{limit2})
one gets \cite{nota2}:
\begin{equation}
\label{gamma}
\Gamma_{_Q}=v_o^4 Q^2 \frac{\pi M}{K_BT} 
\lim_{\omega\rightarrow 0} S(Q,\omega).
\end{equation}
Similarly:
\begin{equation}
\label{tau}
\tau_{_Q} \approx \frac{v_o^4}{(v_\infty^2-v_o^2)}  
\frac{\pi M}{K_BT} \lim_{\omega\rightarrow 0} S(Q,\omega).
\end{equation}
Considering that $\lim_{\omega\rightarrow 0} S(Q,\omega)$ 
is $Q$-independent (see Fig.~1), these expressions give
account of the observation in the low $Q$ region -where
the $Q$ dependence of $v_o$ and $v_\infty$ is negligible-
that the Brillouin peak broadening is proportional to $Q^2$
and the $\tau$ is $Q$-independent. As a check of internal
consistency, using the value of $A_o$ previously reported,
one finds $\tau$$\approx$0.6 ps and $\Gamma_{_Q}$[cm$^{-1}$]=
1.35$Q^2$[nm$^{-1}$]. The value of $\tau_{_Q}$ 
overestimates the one deduced from
Fig.~3, and this probably due to the rough expression
of $\tau$ as $\Gamma_{_Q}/\Delta_{_Q}^2$. On the contrary,
as shown in the inset of Fig.~2,
$\Gamma_{_Q}$ from Eq.~(\ref{gamma}) is in excellent
agreement with the one directly derived from the width
of the $S(Q,\omega)$.

It is worth to discuss the microscopic origin 
of the observed relaxation process. A relaxation process 
can be pictured as the macroscopic manifestation of 
microscopic phenomena associated with the existence of 
channels by which the energy stored in a given "mode" relaxes 
towards other degrees of freedom.
The $S(Q^*,\omega)$, through the fluctuation-dissipation 
relation, reflects the time evolution of the energy initially 
stored ($t$=$t_o$) in a Plane Wave (PW) of wavelength $2\pi/Q^*$. 
As the PW is not an eigenstate of the disordered system, 
at $t\!>\!t_o$ there will be a transfer of amplitude from this 
PW towards other PWs of different $Q$ values. 
This process is controlled by the difference bewteen
the considered PW and the normal modes of the topologically 
disordered glassy structure. This energy flow takes
place on the time scale $\tau$ as derived from Eq.~(\ref{tau}),
and gives rise to the observed relaxation process phenomenology.
Consequently, one can speculate that this process is the
{\it instantaneous} or {\it microscopic} process 
empirically introduced to explain the
$S(Q,\omega)$ measured in real glasses and liquids
by Brillouin light and x-ray Scattering \cite{cumm,IXS}. 

We thanks W.~G\"otze for helpful discussions and 
critical readings of the manuscript.

\newpage

\begin{figure}[f]
\caption[short fig description]{ \footnotesize{
The quantity $(M/K_BT)S(Q,\omega)$ is reported in absolute
value at the indicated $Q$ values for a) the very low $Q$
region ($N$=32000, MD calculation)
and b) the intermediate $Q$ region ($N$=2048, NMA calculation).
The inset of a) shows the comparison ($N$=2048) of the 
MD $S(Q,\omega)$ (full line) and the NMA 
$S^{^{(1)}}\!(Q,\omega)$ ($\circ$).
The inset of b) shows the $Q$ dependence of $M^{(2)}$, the
second moment of $(M/K_BT)S(Q,\omega)$ (+) and its
theoretical prediction $M^{(2)}_{th}$=$Q^2$ (full line).
}}
\label{fig1}
\end{figure} 
\noindent

\begin{figure}[f]
\caption[short fig description]{ \footnotesize{
Frequency position of the maxima of the longitudinal current
spectra ($\Omega_M(Q)$) for different indicated size samples.
The inset shows in log-log scale the maxima ($\Omega(Q)$) 
and the broadening ($\Gamma(Q)$) of the Brillouin peaks
in the low $Q$ region. The full line is the prediction
according to Eq.~(\ref{gamma}) and the dashed line is
the best fit to the data.}}
\label{fig2}
\end{figure} 
\noindent

\begin{figure}[f]
\caption[short fig description]{ \footnotesize{
Sound velocity in glassy LJ Argon: apparent sound velocity ($v$,
full symbols), zero frequency sound velocity ($v_o$, dashed line) 
and ininite frequency sound velocity ($v_\infty$, full line
and open symbols). 
}}
\label{fig3}
\end{figure} 
\noindent


\begin{references}
\bibitem{notatau}
This "relaxation time" $\tau_{_Q}$ can be interpreted as 
the time needed by a set of harmonic oscillators of 
different frequencies to dephase each others.
\bibitem{nota2}
Eq.~(\ref{gamma}) can also be derived from the Fourier
transform of Eq.~(\ref{lang}), i.~e. 
$S(Q,\omega)$= $S(Q)\tilde \phi(Q,\omega)$= 
$S(Q) \omega^2_{o\mu}(Q)/\pi \omega Im 
\left[ \omega^2-\omega_{o\mu}^2(Q)-i\omega\tilde 
m(Q,\omega) \right]^{-1}$, 
performing the $\omega$$\rightarrow$0 
limit, and remembering that the area of the 
memory function is $\tilde m'(Q,\omega=0)$.
\end{references}
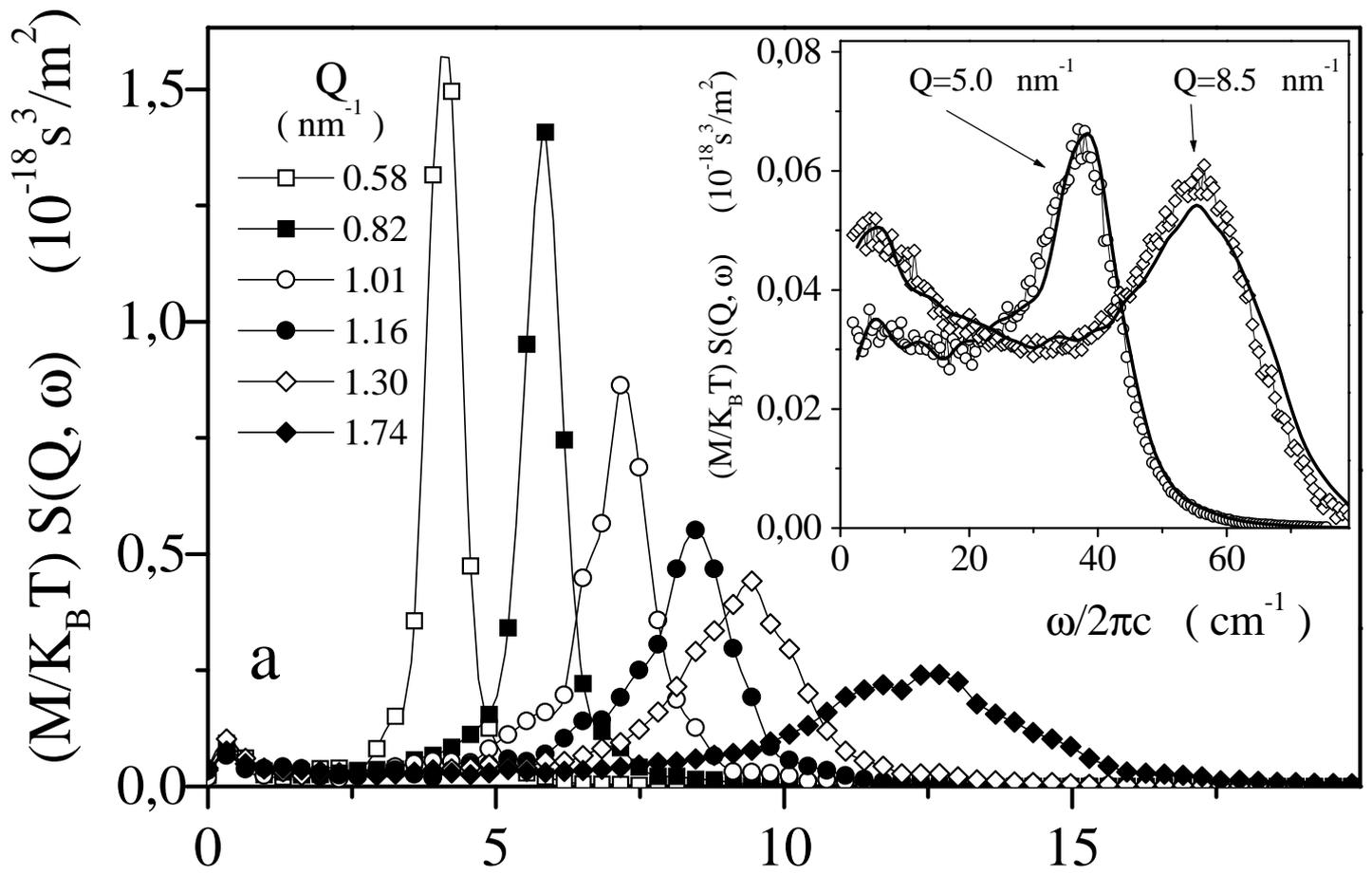
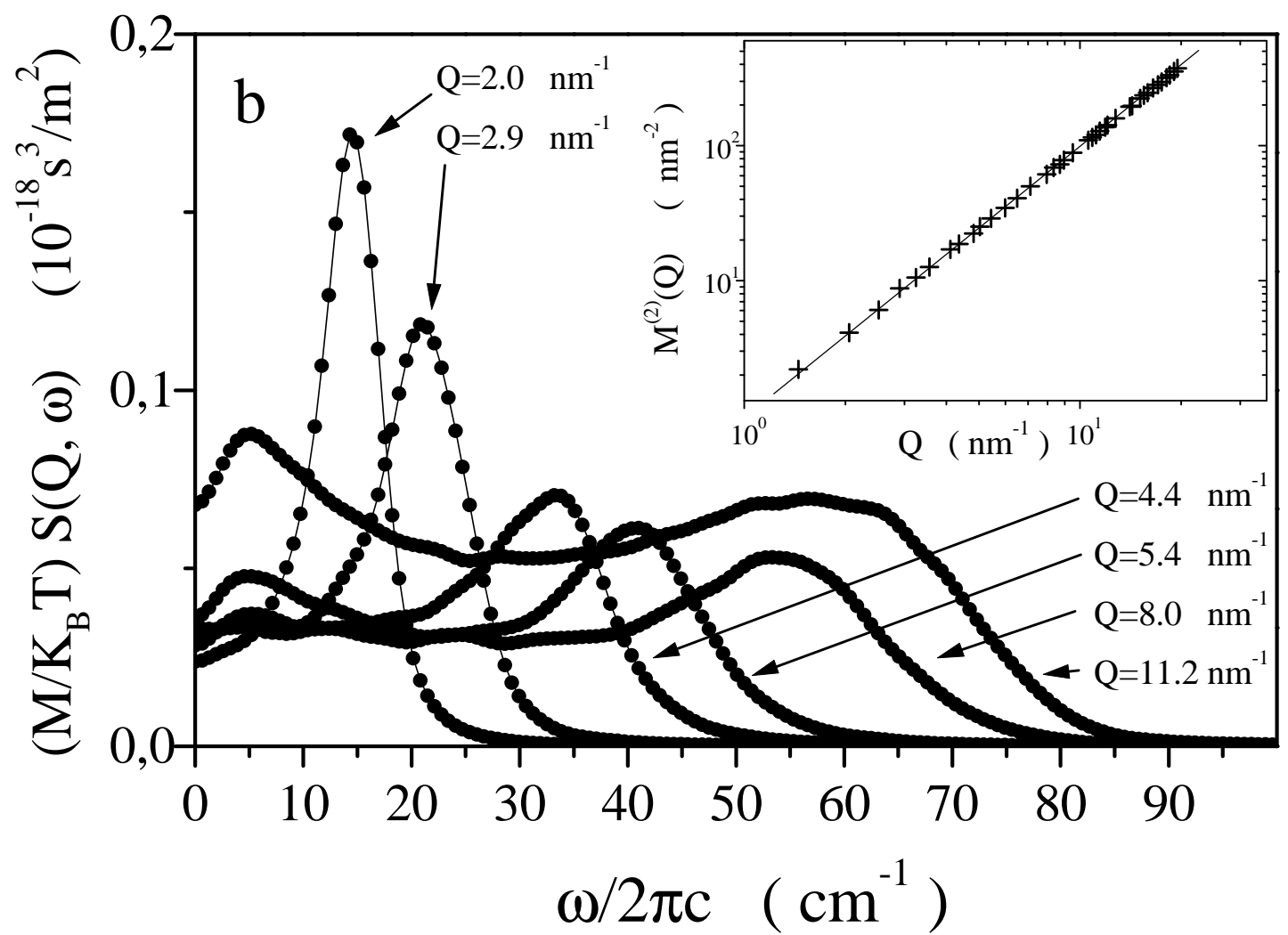
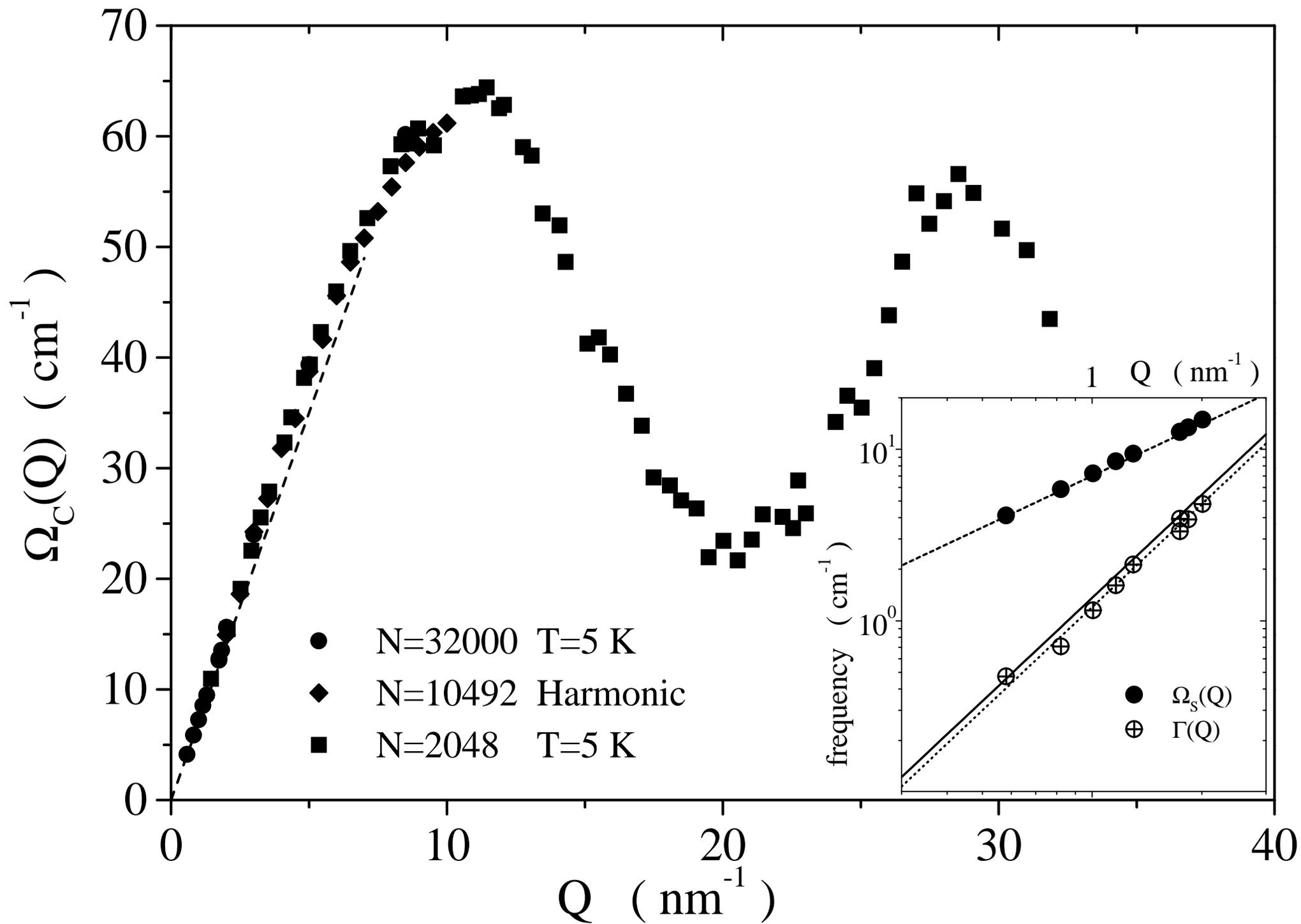
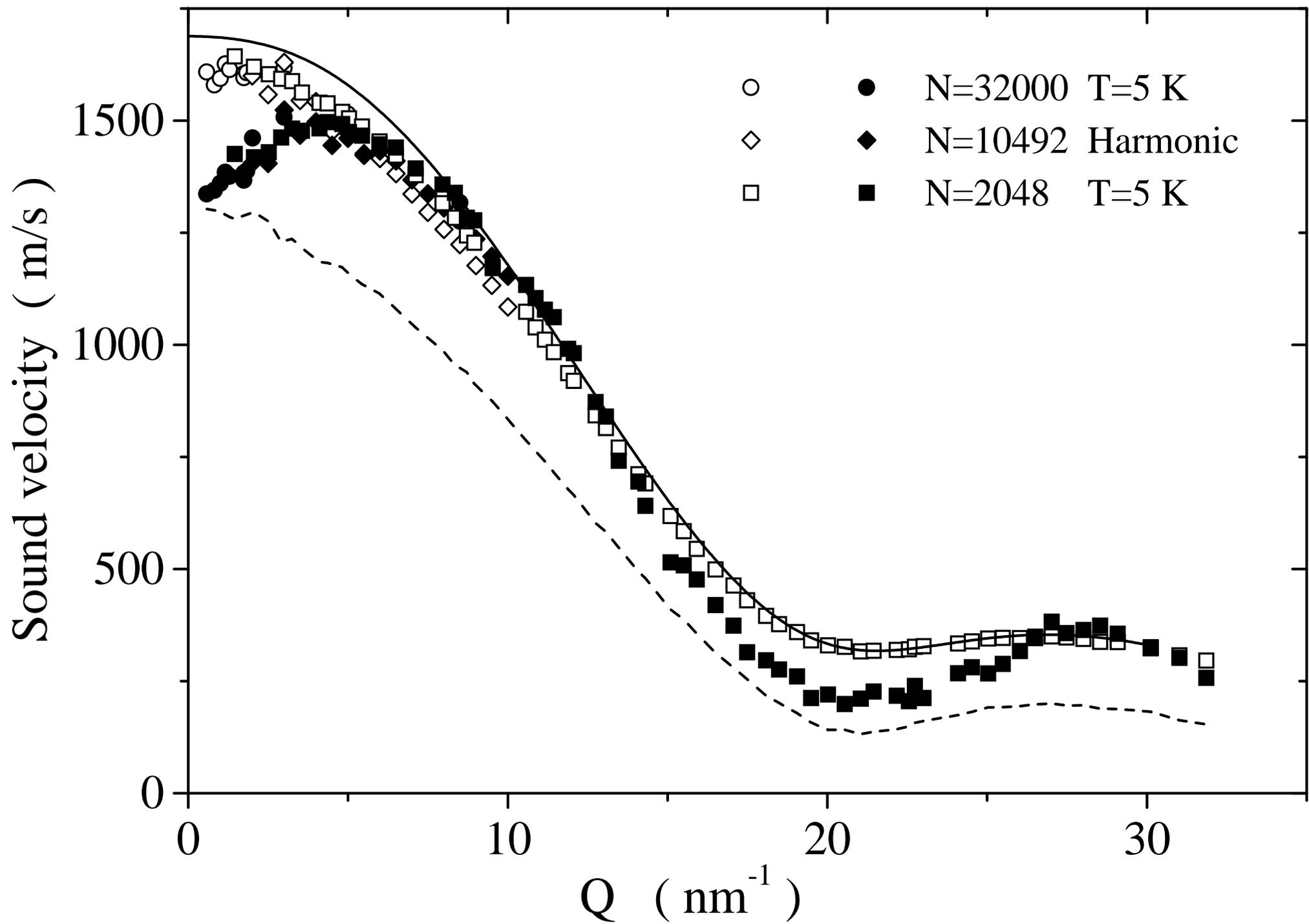
\end{document}